\documentclass{article}

\usepackage[top=2cm,bottom=2cm,right=2cm,left=2cm]{geometry}
\usepackage[utf8]{inputenc}
\usepackage{graphicx} 
\usepackage[usenames, dvipsnames]{color}  
\usepackage[usenames, dvipsnames]{xcolor}  
\usepackage{soul}
\usepackage{xparse}
\usepackage{siunitx}
\usepackage{hyperref}
\usepackage{amsmath}
\usepackage{amssymb}
\usepackage{multirow}
\usepackage{multicol}
\usepackage{printlen}
\usepackage{makecell}
\usepackage{xspace}
\usepackage[skins]{tcolorbox}
\tcbuselibrary{breakable}
\usepackage[switch]{lineno}
\usepackage{sectsty}
\usepackage{authblk}
\usepackage[toc,page]{appendix}
\usepackage{csquotes}
\usepackage{cite}

\newcommand{\lvk}{LIGO-Virgo-KAGRA collaboration}
\newcommand{\pastro}{\ensuremath{p_\text{astro}}\xspace}
\newcommand{\pbns}{\ensuremath{p_\text{BNS}}\xspace}
\newcommand{\pbbh}{\ensuremath{p_\text{BBH}}\xspace}
\newcommand{\pnsbh}{\ensuremath{p_\text{NSBH}}\xspace}
\newcommand{\pterr}{\ensuremath{p_\text{terr}}\xspace}
\newcommand{\psource}{\ensuremath{p_\text{source}}\xspace}
\newcommand{\Msun}{\ensuremath{\mathrm{M_{\odot}}}\xspace}
\newcommand{\mchirp}{\ensuremath{m_{\text{chirp}}}\xspace}
\newcommand{\mtot}{\ensuremath{m_{\text{tot}}}\xspace}
\newcommand{\mchirptot}{\ensuremath{m_{\text{chirp/tot}}}\xspace}
\newcommand{\spineff}{\ensuremath{\chi_\mathrm{eff}}\xspace}

\newcommand{\snrexcess}{\ensuremath{\mathrm{SNR}\mbox{-}\mathrm{Excess}}\xspace}

\title{The MBTA Pipeline for Detecting Compact Binary Coalescences in the Fourth LIGO-Virgo-KAGRA Observing Run}
\author[1]{Christopher Alléné}
\author[2]{Florian Aubin}
\author[3]{Inès Bentara}
\author[1]{Damir Buskulic}
\author[4,5]{Gianluca M Guidi}
\author[6]{Vincent Juste}
\author[3]{Morgan Lethuillier}
\author[1]{Frédérique Marion}
\author[4,5]{Lorenzo Mobilia}
\author[2]{Benoît Mours}
\author[3]{Amazigh Ouzriat}
\author[2]{Thomas Sainrat\footnote{Corresponding author: \href{mailto:thomas.sainrat@iphc.cnrs.fr}{thomas.sainrat@iphc.cnrs.fr}}}
\author[3]{Viola Sordini}
\affil[1]{Univ. Savoie Mont Blanc, CNRS, Laboratoire d’Annecy de Physique des Particules - IN2P3, F-74940 Annecy, France}
\affil[2]{Université de Strasbourg, CNRS, IPHC UMR 7178, F-67000 Strasbourg, France}
\affil[3]{Institut de Physique des 2 Infinis de Lyon (IP2I) - UMR 5822, Université de Lyon, Université Claude Bernard, CNRS, F-69622 Villeurbanne, France}
\affil[4]{Università degli Studi di Urbino ’Carlo Bo’, I-61029 Urbino, Italy}
\affil[5]{INFN Sezione di Firenze, Sesto Fiorentino, I-50019 , Firenze, Italy}
\affil[6]{Service de Physique Théorique, Université Libre de Bruxelles (ULB), Boulevard du Triomphe, CP225, B-1050 Brussels, Belgium}

\date{}

\begin{document}

\twocolumn
\maketitle
\begin{abstract}
    In this paper, we describe the Multi-Band Template Analysis (MBTA) search pipeline dedicated to the detection of compact binary coalescence (CBC) gravitational wave signals from the data obtained by the \lvk~(LVK) during the fourth observing run (O4), which started in May 2023. We give details on the configuration of the pipeline and its evolution compared to the third observing run (O3). We focus here on the configuration used for the offline results of the first part of the run (O4a), which are part of the GWTC-4 catalog (in preparation). We also give a brief summary of the online configuration and highlight some of the changes implemented or considered for the second part of O4 (O4b).
\end{abstract}

\section{Introduction}
The Multi-Band Template Analysis (MBTA) pipeline \cite{MBTA2016,MbtaO3} is one of the several pipelines used by the \lvk~ to search for compact binary coalescences (CBC), along with GstLAL~\cite{GstLAL1,GstLAL2,GstLAL3,GstLAL4}, PyCBC~\cite{PyCBC} and SPIIR~\cite{SPIIR}, as well as the unmodelled search pipeline cWB~\cite{CWB}. It has been reliably operating since the first generation of detectors (LIGO and Virgo science runs S6 and VSR2/VSR3), and has participated  in a number of detections during the past runs of the second-generation detectors~\cite{LIGO,Virgo,KAGRA}. The pipeline is operated both during the run to produce low-latency alerts (\enquote{online}) and after the run to provide updated results for the catalogs (\enquote{offline}) .

MBTA analyzes the data from each detector individually, using a matched-filtering technique, before making coincidences between the candidate events. Its specificity is that the matched-filtering process is split across two frequency bands, reducing the computational cost. It also implements a number of methods for rejecting noise transients (usually referred to as \enquote{glitches}).

In order to improve the sensitivity of the search, and to broaden the capabilities of the pipeline, a number of upgrades have been made to MBTA with respect to~\cite{MbtaO3}. In this paper, we present the configuration and methods used for the offline analysis of O4a data, also highlighting the differences with the online configuration. Section~\ref{sec:description} describes the general features of the pipeline and the configuration for the main search, whose results are reported in the GWTC-4 catalog, while section~\ref{sec:SSM} focuses on the specifics of the search for sub-solar mass objects. Section~\ref{sec:early-warning} presents the online early-warning search, which can produce triggers without waiting for the binary merger.

\section{Description of the pipeline}
\label{sec:description}
In this section, we describe the pipeline as it was used for the O4a offline main search, with subsection~\ref{sec:online} highlighting the configuration specific to the online search. Throughout this section, we illustrate the different parts of the pipeline using results obtained by running the offline configuration on a 40-day period from O3.

\subsection{Preprocessing}
MBTA analyzes the reconstructed strain data provided by the detectors after they have been properly calibrated~\cite{LIGOCalibration,VirgoCalibrationO3} and CAT1 data-quality flags, indicating severe problems in the data, have been made available~\cite{LIGODetchar,VirgoDetCharO3}. 

As in the previous run, MBTA applies several preprocessing steps before the matched-filtering process. The data from the detector is first resampled, usually at \qty{4096}{\hertz} (after applying a low-pass filter). We then apply a gating procedure, which removes stretches of bad-quality data (\enquote{glitches}). The gating is applied when a CAT1 flag is active, or when a short-timescale estimate of the detector sensitivity falls below some threshold; the data is then set to 0, using Tukey windows with 0.5-second taper time to ensure a smooth transition\footnote{This value changed from O3 to cope with the larger dynamics in the spectrum of the LIGO Livingston data in O4.}. The local estimate of the detector sensitivity is based on the so-called binary neutron star (BNS) range, measuring the capacity of the detector to detect fiducial BNS signals, here computed at a rate of \qty{32}{\hertz}. It is obtained by performing a 0.25-second fast Fourier transform (FFT) of the data to estimate the noise power spectral density (PSD); in order to ensure a good accuracy, at each frequency point we take the maximum between the values in this FFT and the median of all the FFTs over the past \qty{10}{\second}. We then apply the gating if the range drops below an adaptive threshold set to $60\%$ of its median value over the last \qty{10}{\second}. Over the O4a run, less than 0.1\% of the observing time was gated by the adaptative threshold for each detector, with gated segments $\sim\qty{1.4}{\second}$-long on average. It should be noted here that short and loud astrophysical signals can trigger the gating; we therefore run an additional ungated search for this type of signals, with tighter constraints than the main search to mitigate the impact of ungated glitches.

In addition, an estimate of the PSD is produced to be used in the matched-filtering process. It is computed from the preprocessed data, by applying FFTs and taking the median of these FFTs on a certain period of time in order to get a smooth estimate; the specifics depend on the configuration and the region of the parameter space, ranging from seconds to hundreds of seconds regarding the length of the FFTs, and thousands of seconds regarding the time over which the median is estimated.
\subsection{Template bank}
\label{sec:templatebank}
The matched-filtering process requires simulated waveforms for the CBC signals, which are commonly called \emph{templates}, and arranged in a template bank to cover a wide parameter space. As in previous searches, we consider this parameter space to be determined by the masses of the objects $(m_1,m_2)$ and their dimensionless spins, assumed to be parallel to the orbital angular momentum $(\chi^z_1,\chi^z_2)$ -- it has been shown that this assumption does not hamper much the detection for the targeted population~\cite{LIGOS3,VanDenBroeck2009,Dal_Canton_2015,GW150914}. Hereafter, we will use the notations \begin{align*}
    \mchirp &= \frac{(m_1m_2)^\frac{3}{5}}{(m_1+m_2)^\frac{1}{5}} & \text{for the chirp mass}\\
    \mtot &= m_1 + m_2 & \text{for the total mass}\\ 
    q &= \frac{m_1}{m_2} \text{ with } m_1 > m_2 & \text{for the mass ratio}   \\
    \spineff &= \frac{m_1\chi^z_1+m_2\chi^z_2}{m_1+m_2} & \text{for the effective spin.}
\end{align*}

Compared to O3, the parameter space has been extended to cover total masses up to \qty{500}{\Msun} in detector frame (\qty{200}{\Msun} in O3), however limiting the mass ratio to 50 (100 in O3). Spin ranges are unchanged from O3, following the astrophysical expectations for merging neutron stars ($|\chi_{1,2}^z| < 0.05$) and black holes ($|\chi_{1,2}^z| < 0.997$); the cutoff between the two is conservatively placed at $\qty{2}{\Msun}$. We also introduce a new lower limit of \qty{200}{\milli\second} on the duration of the templates (starting at \qty{18}{\hertz}), which allows to remove very short templates which tend to be more susceptible to glitches. For the same reason, we remove $67$ templates identified by the online O4a analysis, which have high, asymmetrical masses and high negative effective spin.

Building a template bank requires making a compromise between maximizing signal recovery and minimizing the number of templates. The technical choices and exact parameters values discussed below are the result of such process.
In order to generate the bank, we first consider two regions within the parameter space, corresponding respectively to BNS and BBH with low mass ratio $q$, and generate \enquote{seed} banks for each of them, which are then used as a starting point to build the full bank. This allows us to tightly control the minimal match \cite{Owen_1996} in these specific regions of the parameter space of specific interest for multi-messenger astronomy, and where most of the past events have been found.
A geometric placement algorithm~\cite{Brown:2012qf,Harry:2013tca} is used in O4a to generate the BNS seed bank, ensuring a minimal match of 0.98. This is changed in O4b, where we instead use a stochastic algorithm~\cite{Harry:2009ea} for technical reasons. For the low-$q$ BBH, we use a hybrid algorithm~\cite{hybridalgorithm}\footnote{The geometrical approach relies on the knowledge of the metric, which is reliable only for low masses; for higher masses a hybrid algorithm, which relies both on the metric and a stochastic approach, gives better results.}, with a minimal match of 0.98, as most O3 events fall into that region. Finally, the bank for the full parameter space is initialized using the two seed banks and completed using the hybrid algorithm, with a minimal match of 0.965. The reference PSD used during the generation process is the \enquote{O4 high} sensitivity for the LIGO detectors, with BNS range 190 Mpc\footnote{\url{https://dcc.ligo.org/LIGO-T2000012-v1/public}}. The template models used to generate the bank are the TaylorF2 approximant~\cite{Taylor1,Taylor2} for BNS, while SEOBNRv4~\cite{SEOBNRv4,SEOBNRv4_opt} is used for BBH; we use the implementation from the LALSimulation library~\cite{lalsuite} in both cases. All parameters are summarized in Table~\ref{table:O4UberBankParam}.

\begin{figure*}
    \centering
    \includegraphics[width=\textwidth]{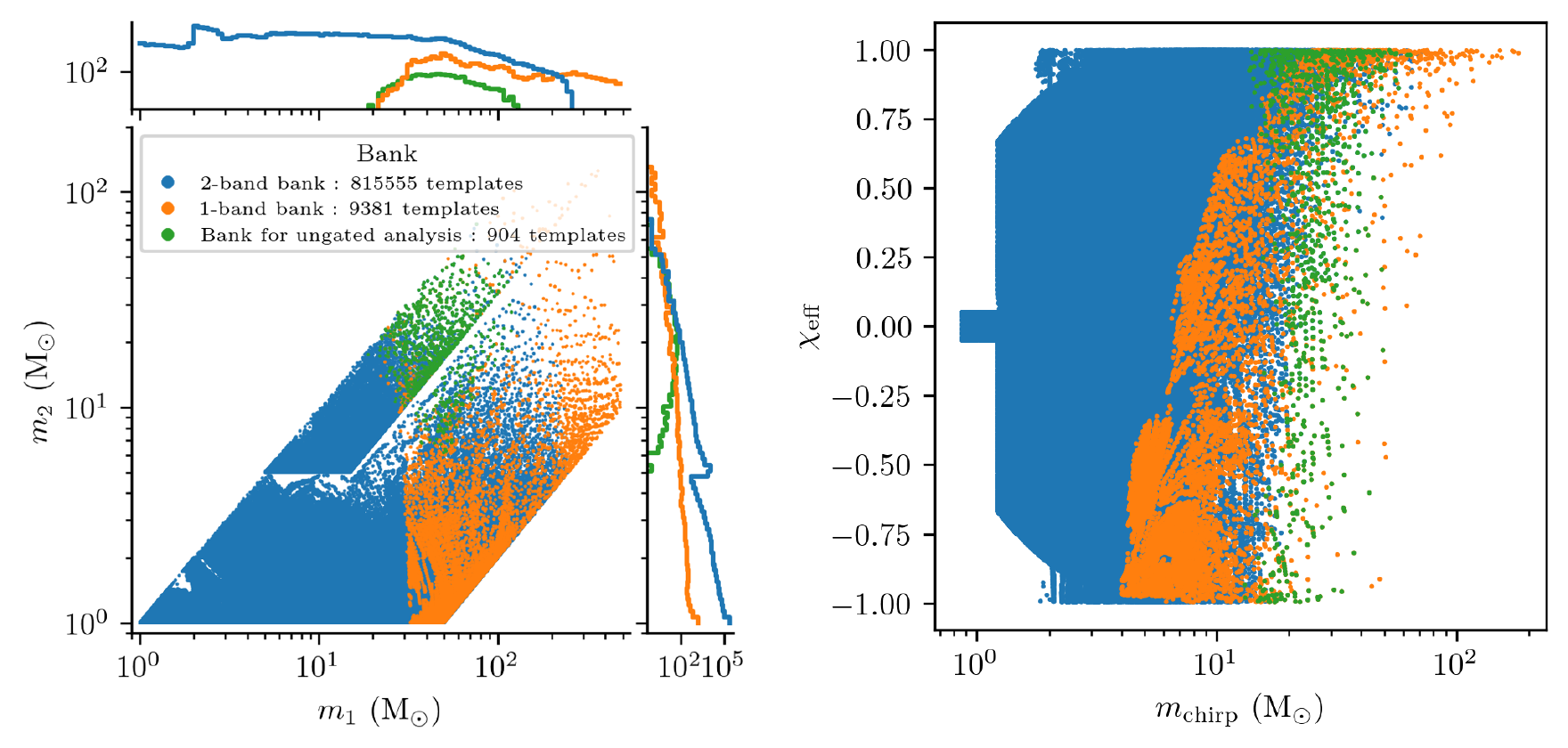}
    \caption{Distribution of the O4a bank templates in the $(m_1,m_2)$ and $(\mchirp,\spineff)$ parameter space. The peculiar form of the bank in the $(m_1,m_2)$ space is explained by the use of seeds. The apparent holes naturally appear when building the bank, and are not a concern as events in these areas are still recovered with other templates. The three colors represent the different sub-banks defined in Section \ref{sec:templatebank}.}
    \label{fig:uberbank}
\end{figure*}

MBTA needs three different template banks: the \enquote{virtual template} bank (containing around 825k templates, shown in Figure~\ref{fig:uberbank}), covering the full frequency range, and two \enquote{real template} banks covering narrower frequency bands, with low-frequency templates (around 55k)  up to \qty{80}{\hertz} and high-frequency templates above (around 20k), with which the matched-filtering process will actually be done. Each virtual template is then associated with a real template from each frequency band, by maximizing their match using a modified version of the PyCBC \texttt{banksim} algorithm\footnote{https://pycbc.org/pycbc/latest/html/banksim.html}. In addition, we set apart the virtual templates merging before \qty{80}{\hertz} to form a \enquote{1-band} bank (around 9k templates). We also select a small (around 1k) sample of templates for the ungated search mentioned in the previous section.

All the banks described here are validated using the \texttt{banksim} algorithm, which is based on Monte-Carlo simulations. These show that more than $97\%$ of the signals in the considered parameter space are recovered with a signal-to-noise ratio (SNR) loss lower than $3\%$ ($1.5\%$ loss on average).

\begin{table*}
    \centering
    \begin{tabular}{ ||c|c|c|c|c|| } 
    \hline
     & \multirow{2}{*}{BNS seed} & \multirow{2}{*}{BBH seed} & Full bank & \multirow{2}{*}{Subsolar mass bank} \\
     &  &  & (seeds included) & \\ 
    \hline
    \hline
    Number of virtual templates & 218182 & 37208 & 825840 & 2253561 \\\hline
    Individual masses ($m_{1,2}$) [\Msun] & $\left[1, 3\right]$ & $\left[5, 500\right]$ & $\left[1, 500\right]$ & $\left[0.2, 10.0\right]$ \\ 
    Total mass ($m_1 + m_2$) [\Msun] & $\left[2, 4\right]$ & $\left[10, 500\right]$ & $\left[1, 500\right]$ & $\left[0.2, 1.0\right]$ \\ 
    Mass ratio ($m_1 / m_2 \geq 1$) & $\left[1, 3\right]$ & $\left[1, 3\right]$ & $\left[1, 50\right]$ & $\left[1.0, 10.0\right]$ \\ 
    \multirow{2}{*}{Individual aligned spins ($\chi^z_{1,2}$)} & \multicolumn{3}{c|}{$\left[-0.05, 0.05\right]$ if $m_{1,2} \leq 2 ~\Msun$} & $\left[-0.1,0.1\right]$ if $m \leq 0.5 \Msun$ \\ 
     & \multicolumn{3}{c|}{$\left[-0.997, 0.997\right]$ else } & $\left[-0.9, 0.9\right]$ else \\
    \hline
    Minimal duration from \qty{18}{\hertz} [s] & - & \multicolumn{2}{c|}{$0.2$} & - \\
    \hline
    Minimal match & \multicolumn{2}{c|}{$0.98$} & $0.965$ & 0.97 \\
    \hline
    Approximants for bank generation & TaylorF2 & \multicolumn{2}{c|}{SEOBNRv4\_ROM} & TaylorF2 \\
     Approximants for analysis & SpinTaylorT4 & \multicolumn{2}{c|}{SEOBNRv4\_opt} & SpinTaylorT4 \\
    \hline
    PSD for bank generation & \multicolumn{4}{c||}{L1 O4 high-sensitivity (BNS range $\sim 190 ~\mathrm{Mpc}$)} \\
    \hline
    $f_0$ [Hz] for bank generation & $25$ & \multicolumn{2}{c|}{$18$} & $45$ \\
    $f_0$ [Hz] for analysis & $24$ & \multicolumn{2}{c|}{$20$} & $45$ \\
    \hline
    $f_{max}$ [Hz] for bank generation & $2048$ & \multicolumn{2}{c|}{$1024$} & $1000$ \\
    $f_{max}$ [Hz] for analysis  & \multicolumn{3}{c|}{$2048$} & $1000$ \\
    \hline
    $f_{c}$ [Hz] & \multicolumn{3}{c|}{$80$} & $120$ \\
    \hline
    \end{tabular}
    \caption{Parameters of the template banks used for the O4a MBTA main search and sub-solar mass search. $f_0$ represents the starting frequency for the templates, while $f_{max}$ is the maximum frequency. $f_c$ corresponds to the frequency cutoff between the two frequency bands. Also note that the waveform approximants used for bank generation are all in the frequency domain, while the ones for the analysis are in the time domain. Differences for the O4b bank are highlighted in Section \ref{sec:online}.}
    \label{table:O4UberBankParam}
\end{table*}

\subsection{Triggers and coincidences}
\label{sec:triggers}
For each detector, MBTA applies the matched-filtering process with every real template and the results from the real templates are summed coherently by applying a time translation and a phase rotation to yield the SNR as a function of time for each virtual template. We check the maximum SNR over a typical timescale of \qty{4}{\second} against a threshold $\rho_\text{min}$ (set to 4.4 for the main search and 10 for the ungated search) and record a trigger if this value is exceeded. At this stage, we also apply a basic $\chi^2$ cut, where we compare the distributions of SNR in the two frequency bands to that expected from an astrophysical signal, as we describe in~\cite{MBTA2016}.
As in O3, we then define an initial ranking statistic (RS), given by the SNR reweighted with a $\chi^2$ statistic comparing the SNR timeseries with the template autocorrelation, or \enquote{auto$\chi^2$}. We use the same formula and parameters given in~\cite{MbtaO3}. At this stage, we also apply the data-quality reweighting described in section~\ref{sec:snrexcess}.

In order to identify candidate events, we look for coincident triggers between two detectors sharing the same template within a conservative time of flight window between the two detectors (for instance \qty{15}{\milli\second} between LIGO Hanford and LIGO Livingston). An O4 novelty is that we also consider some single-detector triggers on their own, as described in section~\ref{sec:singles}.

We compute the ranking statistic for a coincidence by summing quadratically the ranking statistics of the single detector triggers, and adding a term measuring the consistency of arrival times and phases\footnote{In contrast with O3, we do not include the relative amplitude term due to its minor impact.} across the detectors: \begin{equation}\text{cRS}_{ij}^2 = \text{RS}^2_i + \text{RS}^2_j + 2 \ln(P_{ij})
\label{eq:cRS}
\end{equation} where $\text{RS}_i$ is the ranking statistic for detector $i$ and $P_{ij}$ depends on the arrival time and phase differences between the detectors; the latter are compared to the expected distributions assuming a uniform source population, as described in~\cite{MbtaO3}.
At low ranking statistics, the trigger distribution is mostly populated by noise. For further analysis, we only consider triggers with a ranking statistic higher than 7.

In order to improve the sky localization~\cite{FujiiSkyLocalization,UserGuide}, we also include SNR timeseries from detectors which did not produce a trigger; however, the ranking statistic is left unchanged\footnote{In~\cite{MbtaO3}, we described a method for computing a ranking statistic for three-detector triggers, but it was not used during O4.}. This is for instance used during O4b when Virgo data is available. 

We cluster triggers that are close in time (\qty{100}{\milli\second} from any other trigger of the cluster), retaining the trigger with the highest ranking statistic as representative of the cluster.

\subsection{Internal data quality assessment}
\label{sec:snrexcess}
\begin{figure}[ht]
    \centering
    \includegraphics[width=\columnwidth]{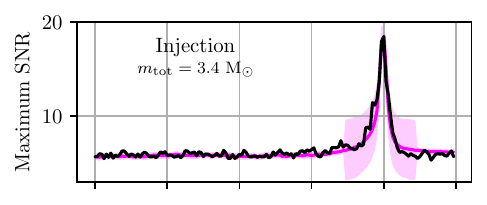}
    \includegraphics[width=\columnwidth]{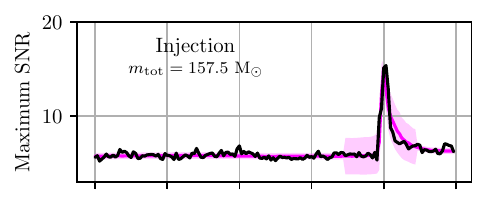}
    \includegraphics[width=\columnwidth]{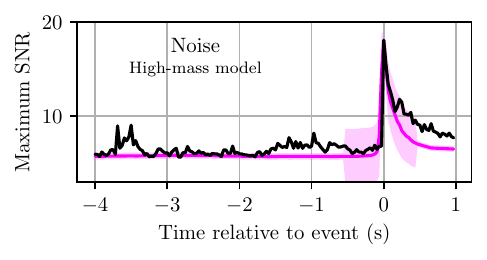}
    \caption{Illustration of the SNR-Excess process. In each case, the black curve corresponds to the timeseries of the SNR maximized over the template bank, while the magenta one is the model (the shaded region represents the uncertainty). The top two figures correspond to well behaved injected signals matching the two different models, with $\chi^2 < 1$; the bottom one corresponds to a glitch within a noisy period, with $\chi^2 \simeq 7$ reducing the ranking statistic by $\sim 30\%$.}
    \label{fig:snrexcess}
\end{figure}
During the O3 run, we introduced a new observable $E_R$, penalizing triggers that occur during periods showing an excess in the trigger rate, signaling poor data quality (see section 4 of~\cite{MbtaO3}).
This procedure was used until the end of the online O4a analysis.
Although it was able to reject a number of transient noises, it showed limitations in identifying short noisy periods or glitches impacting the less dense parts of the bank (as those do not result in a significant trigger rate increase), as illustrated e.g. by the low-latency alert S230622ba~\cite{S230622ba}, which had to be retracted.

\begin{figure}
    \centering
    \includegraphics{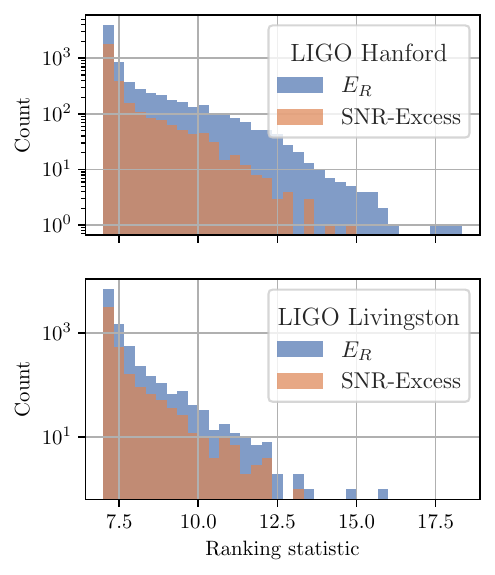}
    \caption{Distribution of the ranking statistic of single-detector triggers from the 40-day O3 rerun when applying the \snrexcess (orange) and $E_R$ (blue) SNR reweighting.}
    \label{fig:snrexcessperformance}
\end{figure}

We thus decided to replace it with a new technique called \enquote{SNR-Excess}, still applied in post-processing as it uses the full template bank information.
It takes the timeseries of the maximum SNR over the template bank\footnote{Whereas the $\mathrm{auto}\chi^2$ only relies on the SNR timeseries for a specific template.} (for a period of $-4$ and $+1$ seconds around a given trigger, with a \qty{32}{\hertz} sampling rate) and compares it \textit{via} a $\chi^2$ statistic to a model based on a large population of simulated signals (injections) in O4a data covering the whole parameter space. We expect that an excess of noise may trigger many templates in the bank, while signals should only trigger matching templates, which usually allows to get a different response.
As the pipeline responds differently based on detected parameters, two models are used: one for triggers with (detector-frame) total mass below $25~\mathrm{M_{\odot}}$ and another above. Figure~\ref{fig:snrexcess} illustrates this comparison in the cases of injections and of a noise trigger.
The $\chi^2$ is renormalized to eliminate its SNR dependency. The \snrexcess $\chi^2$ is used to reweight the single detector ranking statistic (RS), which is redefined as
\begin{equation}
    \mathrm{RS} =
    \left\{
        \begin{array}{ll}
            \mathrm{RS} & \mbox{if } \chi_{\snrexcess}^2 < 1 \\
            \mathrm{RS} \times \left( \frac{1}{\chi_{\snrexcess}^2} \right)^{0.2} & \mbox{else.}
        \end{array}
    \right.
    \label{eq:snrexcessreweighting}
\end{equation}
Figure~\ref{fig:snrexcessperformance} displays the ranking statistic distribution when using the \snrexcess or the $E_R$.

In addition, we reject part of the coincidences for which in either one of the detectors the ranking statistic was decreased by more than 15\% after applying the auto$\chi^2$ and the \snrexcess. This cut is only applied to the noisiest part of the parameter space ($m_\mathrm{tot} > \qty{106}{\Msun}$ for $\spineff > 0.33$, $m_\mathrm{tot} > \qty{45}{\Msun}$ for  $-0.33 < \spineff < 0.33$ and $m_\mathrm{tot} > \qty{34}{\Msun}$ for $\spineff < -0.33$), and not applied to coincidences having an SNR above 9 in both LIGO detectors.

\subsection{Single-detector events} 
\label{sec:singles}
A new addition for the O4 run is to estimate the significance of triggers from a single detector, which other CBC pipelines have also implemented~\cite{GstLALO2,GstLAL2,PyCBCsingles}. This is motivated by the fact that about 14\%\footnote{Based on the CBC\_CAT2 timelines from GWOSC} of the O3 run time had only one detector online, which may contain a significant number of events occurring during that time; we also expect to recover some events where two detectors are online but there is substantial signal in only one of them, which is susceptible to happen with one of the LIGO detectors and the less sensitive Virgo, representing an additional $\sim 25\%$ of the run time. In O4, we only considered single-detector triggers from LIGO detectors, whether Virgo was operating or not.

In order to compute the background for estimating the significance (see section~\ref{sec:pastro}), we cannot use the usual fake coincidence technique~\cite{MbtaO3} as there is only one detector. Instead, we use the single-detector triggers distribution during double-detector time, removing coincident events to reduce contamination from real events. This distribution is fitted with an exponential function in order to extrapolate to high ranking statistic values. The validity of this extrapolation was verified by designing a different way to do the fake coincidences, based on the multi-band feature of MBTA; these investigations are described in chapter 6 of~\cite{thesevincent}.

\begin{figure}
    \centering
    \includegraphics{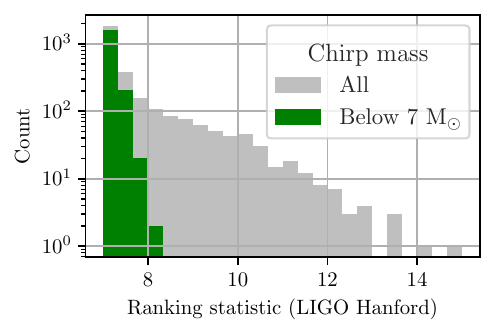}
    \caption{Ranking statistic distribution with and without the single-detector chirp mass threshold at $\qty{7}{\Msun}$.}
    \label{fig:rankingstat_chirpmasscut}
\end{figure}

For the O4a search, we chose to only consider single-detector events with a chirp mass below $7~\Msun$, as illustrated by Figure~\ref{fig:rankingstat_chirpmasscut}. These events cover the BNS and most of the NSBH parameter space, which are the main targets for electromagnetic follow-up; they match longer templates, making them easier to distinguish from noise.

\subsection{Probability of being astrophysical and source classification}
\label{sec:pastro}
A first metric of the significance of a trigger is the probability that it is of astrophysical origin, $\pastro = 1 - \pterr$ (\pterr representing the probability of the trigger being of terrestrial origin).
To compute \pastro, we use a method similar to the one developed for O3~\cite{MBTAO3pastro,GWTC3}. We define \pastro as \begin{equation}
    \label{eq:pastro}
    \pastro(\text{cRS}) = \frac{n_f(\text{cRS})}{n_f(\text{cRS})+n_b(\text{cRS})}
\end{equation}

with $n_f$ and $n_b$\footnote{Related respectively to the $\Lambda_1$ and $\Lambda_0$ rates in \cite{MBTAO3pastro}.} representing the densities over cRS (defined in equations \ref{eq:cRS} and \ref{eq:snrexcessreweighting}) of the expected number of foreground (astrophysical) and background (noise) events.
To take into account the variations of the foreground and background over the parameter space, this equation is evaluated over a number of bins that are defined over the chirp mass (switching to total mass when $\mchirp > \qty{7}{\Msun}$), the mass ratio (switching to effective spin when $\mchirp > \qty{7}{\Msun}$) and the possible detector combinations. The mass ratio and effective spin axes are divided into 3 bins each, while the chirp mass axis is divided in 43 bins, and the total mass in 40 bins; this gives 213 $(\mchirptot, q/\spineff)$ bins containing at least one template.

The background $n_b(\text{cRS})$ is measured on a reference dataset using a similar method to O3~\cite{MbtaO3}. Its logarithm is fitted with third-order polynomials; it is also rescaled by a \enquote{clustering factor} accounting for the reduction of the number of triggers during the clustering step. 

\begin{figure}[h!]
    \centering
    \includegraphics{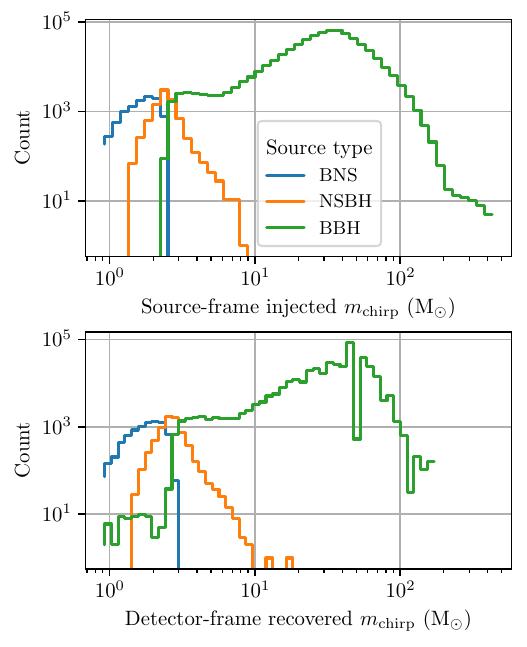}
    \caption{Distribution of the chirp mass (with logarithmic bins) for the various source types used for building the \pastro model. The top part shows the distribution of the injections (in the source frame), while the bottom part shows the events recorded by the pipeline (with a ranking statistic above 8 and within $\pm\qty{0.1}{\second}$ of an injection).}
    \label{fig:foreground_distrib}
\end{figure}

The foreground is estimated by assuming the same population model as the offline injections (adjusted to the number of detections) used by the LVK to evaluate the search sensitivity for the GWTC-4 catalog. The foreground model is built by running the pipeline on these injections, which provides the distribution of the detected masses, taking into account the discreteness of the template bank. Figure~\ref{fig:foreground_distrib} displays the chirp mass distribution of the population model in the source frame and as detected by the pipeline.
In order to obtain the foreground as a function of the cRS, we also assume that the foreground cumulative distribution is proportional to $\text{SNR}^{-3} \approx \text{cRS}^{-3}$.

\begin{figure*}
    \centering
    \includegraphics{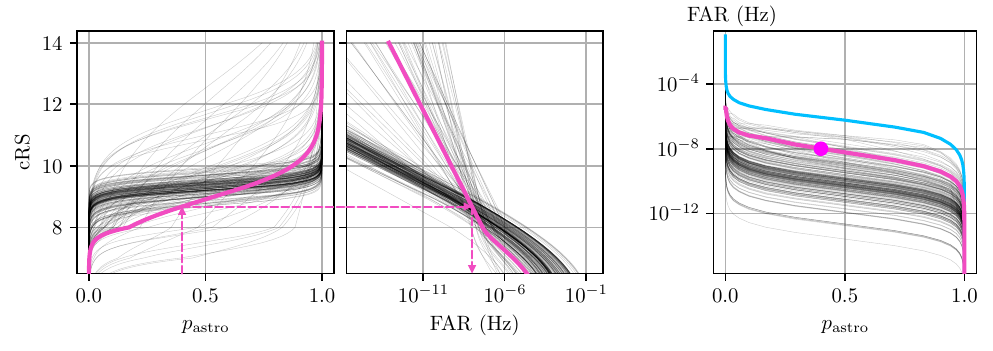}
    \caption{Representation of the process for generating the FAR(\pastro). Each of the black lines represents a different ($\mchirptot$, $q/\spineff$, HL) bin; the magenta line is an example of a bin with high-mass components. The first plot shows the (inverted) $p_{\text{astro}(i,\text{HL})}(\text{cRS})$ relation, while the second one shows the (inverted) $\text{FAR}_{(i,\text{HL})}(\text{cRS})$; a given value of \pastro yields a FAR value as shown by the arrows. The plot on the right shows the resulting $\text{FAR}_{(i,\text{HL})}(\pastro)$, with the dot representing the point determined from the arrows; the blue curve is the sum of all the individual curves, yielding $\text{FAR}_\text{HL}(\pastro)$.}
    \label{fig:pastro_far_process}
\end{figure*}

This process yields a relationship between \pastro and the ranking statistic for each bin, which can be parametrized. This parametrization is then used to obtain the \pastro value of a trigger, both for the online and offline analyses. The left plot of Figure~\ref{fig:pastro_far_process} displays this parametrization for the different bins.

\begin{figure}
    \centering
    \includegraphics{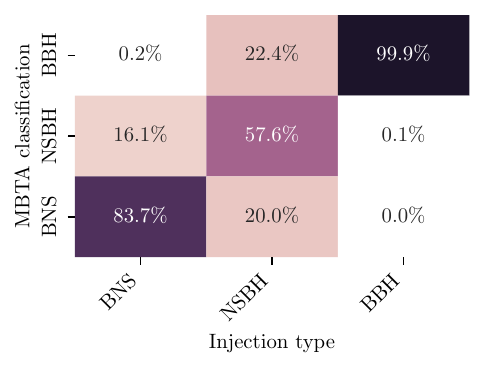}
    \caption{Confusion matrix for the source classification (according to the largest \psource value). Among the NSBH misclassified as BBH, 73.6\% still have $\pnsbh~>~0.1$.}
    \label{fig:confusion_matrix}
\end{figure}

In addition, we split \pastro into three source-type probabilities, such that $\pastro = \pbns + \pnsbh + \pbbh$. These are obtained by multiplying \pastro by the relative rate of the corresponding source type recovered within each bin. In the assumed astrophysical population, the threshold between neutron star and black hole is placed at $\qty{3}{\Msun}$. Figure~\ref{fig:confusion_matrix} shows the confusion matrix between the MBTA classification and the true source type of the corresponding injections.

\subsection{False alarm rate}
\label{sec:FAR}
In O3, the false alarm rate (FAR) was computed using a ranking statistic similar to the one described in section~\ref{sec:triggers}. In O4, in order to take into account the non-uniformity of sources across the parameter space (introduced by astrophysical priors), we use the \pastro described in the previous section as our final ranking statistic. We thus build a global FAR(\pastro), by estimating the number of background events above any given \pastro value. This has the additional benefit of ensuring consistency between the FAR and \pastro values.

The first step is done independently for each ($\mchirptot$, $q/\spineff$, detector combination) bin, which we will denote by the $(i,d)$ indices (with $i$ representing the intrinsic parameters and $d$ the detector combination, HL, H-Lon, L-Hon, H and L\footnote{H and L represent LIGO Hanford and Livingston respectively; the \enquote{-on} means that the SNR in the second detector was too low to make a coincidence. During O4b, we do not differentiate triggers with or without Virgo.}). We aim to build a $\mathrm{FAR}_{(i,d)}$(\pastro) parametrization, which is done  by combining the $p_{\text{astro},(i,d)}(\text{cRS})$ and $n_{b,(i,d)}(\text{cRS})$ parametrizations obtained in the previous section. The $p_{\text{astro},(i,d)}(\text{cRS})$ parametrization is inverted to get a $\text{cRS}_{(i,d)}(\pastro)$; the $n_{b,(i,d)}(\text{cRS})$ distribution corresponds to the one used to compute the \enquote{background} term in \pastro. By combining the two, we get a one-to-one relationship between \pastro and the local FAR for a given bin. Figure~\ref{fig:pastro_far_process} illustrates this process.

We then combine the results for different detector combinations to create a single parametrization for each ($\mchirptot,q/\spineff$) bin, which is obtained with
\begin{equation}
    \label{eq:FAR_detectors}
    \mathrm{FAR}_i(\pastro) = \sum_{\substack{d \in \{\mathrm{HL, H, L,}\\ \mathrm{H-Lon, L-Hon}\}}} w_d\ \mathrm{FAR}_{(i,d)}(\pastro)
\end{equation}
where $w_d = T_d / T_\text{total}$ represents the fraction of the observing time for this detector combination over the total time $T_\text{total}$ (with $T_\text{total} = T_\mathrm{HL} + T_\mathrm{H} + T_\mathrm{L}$ and $T_\mathrm{H-Lon} = T_\mathrm{L-Hon} = T_\mathrm{HL}$)

Finally, we combine all the bins together to form a single FAR(\pastro) mapping, allowing to obtain a consistent relationship between FAR and \pastro, by simply computing
\begin{equation}
    \mathrm{FAR}(\pastro) = \sum_{i \in \mathcal{O}} \mathrm{FAR}_i(\pastro)
    \label{eq:FAR_global}
\end{equation}
where $\mathcal{O}$ represents the set of 213 $(\mchirptot,q/\spineff)$ bins.

\begin{figure}
    \centering
    \includegraphics[width=0.5\textwidth]{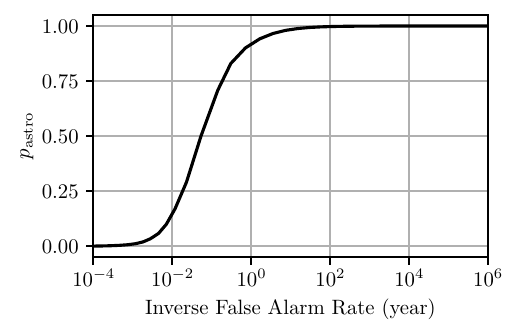}
    \caption{Relation between \pastro and the inverse false alarm rate (IFAR), derived from the FAR(\pastro) parametrization, for the first two months of the O4a offline analysis.}
    \label{fig:farpastro}
\end{figure}

\begin{figure}
    \centering
    \includegraphics{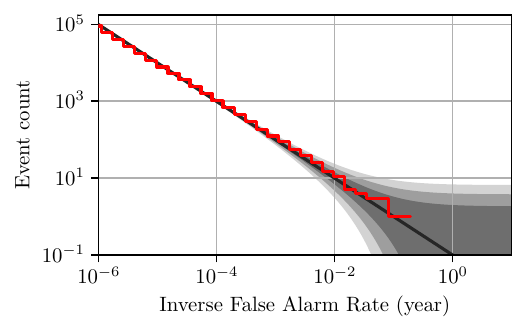}
    \caption{Cumulative distribution of the IFAR of the triggers from the partial O3 rerun. The black line represents the expected amount of noise triggers for 36.6 days of observation, with error bars as the shaded regions. Triggers matching events from GWTC-3 were removed.} 
    \label{fig:ifar_cumulative}
\end{figure}

For the offline run, these models are built for each chunk of 2 months. Figure~\ref{fig:farpastro} illustrates the resulting FAR(\pastro) model. Figure~\ref{fig:ifar_cumulative} shows that, in the noise-dominated regime, the number of events at a given FAR threshold behaves according to the expectation.
For the online run, we instead use a different model for each possible source type, as described in~\ref{sec:farpastroappendix}.

\subsection{Online operations}
\label{sec:online}
The main search has been running online since the beginning of O4. The configuration during O4a had a few differences with the offline analysis: \begin{itemize}
    \item The 67 templates mentioned in subsection~\ref{sec:templatebank} were not removed.
    \item The \snrexcess reranking and cut discussed in section~\ref{sec:snrexcess} were not yet commissioned, and thus the older \enquote{$E_R$} method was used.
    \item The \pastro model was the one used for the O3b offline analysis (see~\cite{MBTAO3pastro}), adjusting the event rates to match the increased sensitivity.
    \item As detailed in~\ref{sec:farpastroappendix}, we used separate FAR(\pastro) models depending on the source type, in order to improve the chance of detecting BNS and NSBH sources.
\end{itemize}

For the O4b online configuration, we integrate the following changes:
\begin{itemize}
    \item A new template bank was generated using noise curves measured during O4a; we also restrict the maximum mass ratio within the bank to less than 20 (instead of 50).
    \item The \snrexcess reranking is part of the online search, without the cut introduced at the end of section \ref{sec:snrexcess}.
    \item The \pastro model is the same as the one for the O4a offline analysis. 
    \item We still use a different FAR(\pastro) model per source type.
    \item We also filter data from Virgo, which joined the second part of the run.  Given the lower sensitivity compared to the two LIGO detectors, the data is not used to estimate the significance, but to improve sky localisation.
\end{itemize}
In Table~\ref{tab:O4a_stats}, we provide a summary of the performance of the various online searches during the O4a run. 

\begin{table*}[h!]
    \centering
    \begin{tabular}{lccccc}
        & & & \multicolumn{3}{c}{Early-warning} \\ \cline{4-6}
        Search & Main  & SSM & \qty{42}{Hz} & \qty{50}{Hz} & \qty{58}{Hz} \\ \hline
        Latency 50\% [5\%;95\%] quantiles (s) & $ 21 ~[19;25]$ & $47 ~[42;55]$ & $-6 ~[-15;7]$ & $6 ~[-3;11]$ & $9 ~[2;19]$ \\
        Low-significance events & 36 & $\sim 50$ & 4 & 4 & 11\\
        \makecell[l]{Low-significance events seen as \\ significant by another pipeline} & 18 & / & \multicolumn{3}{c}{0}\\
        Significant events & 30 & / & \multicolumn{3}{c}{0}\\
        Retractions & 1 & / & \multicolumn{3}{c}{0}
    \end{tabular}
    \caption{Summary of the results of the various online searches during the O4a run, based on the events uploaded to the gravitational-wave events candidate database (GraceDB). An event is labelled as \enquote{low-significance} if its FAR is below $4.63\times10^{-6}\ \unit{\hertz}$ and \enquote{significant} if it is below $7.72\times10^{-8}\ \unit{\hertz}$, corresponding respectively to 2/day and 1/month, adjusted with a trial factor of 5 to account for the various pipelines (these thresholds are the ones used for public alerts, as detailed in \url{https://emfollow.docs.ligo.org/userguide/analysis/index.html}). As a reference, all pipelines produced a total of 1610 low-significance alerts, 81 validated significant alerts (including 8 single-detector events with $\mchirp > \qty{7}{\Msun}$, not targeted by MBTA) and 11 retractions. The latency is given relative to the merger time, and represents the time when the event is successfully registered in GraceDB. This includes around \qty{10}{\second} of latency to receive the data from the detectors. The first number corresponds to the median latency, while the numbers inside brackets are the 5\% and 95\% quantiles. We do not include event counts for the SSM search as no results were made public during O4a. The SSM search is described in Section \ref{sec:SSM}, and the early-warning searches in Section \ref{sec:early-warning}.}
    \label{tab:O4a_stats}
\end{table*}

\subsection{Performance illustration}
As mentioned at the beginning of the section, we ran the O4a offline configuration of the pipeline on a 40-day period from O3. In Table~\ref{table:recovered_events}, we list the most significant events recovered by this test run, which we compare to the events from GWTC-3~\cite{GWTC3}. Out of the 9 GWTC-3 events from this period, 6 are recovered significantly ($\pastro > 0.5$). Among the 3 remaining events, two of them were not very significant in GWTC-3, while the third one is effectively a single-detector trigger in the O4a configuration (as Virgo data is not analyzed), with a chirp mass above the threshold of $\qty{7}{\Msun}$. We also identify an additional candidate with a \pastro slightly above 0.5.

\begin{table*}
    \centering

\begin{tabular}{cclc|clclc}
\multicolumn{4}{c|}{O3 rerun} & \multicolumn{3}{c}{GWTC-3} & \multicolumn{2}{c}{MBTA GWTC-3}  \\\hline
GPS time        & RS  & IFAR (y)     & \pastro & Identifier & IFAR (y) & \pastro & IFAR (y) & \pastro \\ \hline
        $1264316116.4$ & $16.3$ & 9.06e+07 & 1.00 & GW200129\_065458     & 1e+05   & 0.99 &        & \\
        $1264693411.6$ & $10.8$ & 6.11e+06 & 1.00 & GW200202\_154313     & 1e+05   & 0.99 &        & \\
        $1263097407.7$ & $10.9$ & 3.32e+06 & 1.00 & GW200115\_042309     & 1e+05   & 0.99 & 182     & 0.99\\
        $1264213229.9$ & $9.1$ & 4.76    & 0.96 & GW200128\_022011     & 233     & 0.99 & 0.303   & 0.98\\
        $1265202095.9$ & $9.9$ & 2.64    & 0.93 & GW200208\_130117     & 3.23e+03 & 0.99 & 2.17    & 0.99\\
        $1265273710.2$ & $9.0$ & 0.341   & 0.70 & GW200209\_085452     & 21.7    & 0.97 & 0.0833  & 0.97\\
        $1264333383.1$ & $7.8$ & 0.24    & 0.61 &                      &        &  &        & \\ \hline
        $1265235995.9$ & $7.9$ & 0.0291  & 0.14 & GW200208\_222617     & 0.208   & 0.70 & 0.00238 & 0.02\\
        $1265361793.0$ & $8.6$ & 0.00355 & 0.01 & GW200210\_092254     & 0.833   & 0.54 &        & \\
        $1262879936.1$ & $13.6$ & N/A    & N/A & GW200112\_155838     & 1e+05 & 0.99  &     \\
\end{tabular}
    \caption{Most significant events retrieved by the search (\pastro $>$ 0.5) in the partial re-analysis of O3 data. We are able to recover more events than the pipeline contributed to GWTC-3, and with increased IFAR values for MBTA events already in GWTC-3. The improvement depends on the region of the parameter space where the event is recovered. There are 3 GWTC-3 events from the period which were not retrieved : GW200112\_155838, which was only observed by LIGO Livingston and Virgo, and thus seen as a single detector event (with \mchirp around 27) in the O4 configuration, and GW200208\_222617 and GW200210\_092254, with respective FAR values of 4.8 per year and 1.2 per year in the catalog, which were not seen as significant in the O3 MBTA offline analysis either. We also retrieve an additional event which was not included in GWTC-3 as it was below the \pastro $>$ 0.5 threshold; it was however reported in~\cite{4OGC} as GW200129\_114245, with a \pastro of 0.53 and IFAR of \qty{0.04}{y}.}
    \label{table:recovered_events}
\end{table*}

\section{Sub-solar mass search}
\label{sec:SSM}
An interesting target for gravitational-wave astronomy are compact objects with a mass possibly below 1 solar mass (sub-solar mass or SSM objects). There is no widely accepted astrophysical channel that predicts the existence of compact objects in this mass range, but several possibilities have been proposed. These include primordial black holes or dark black holes~\cite{SSMO3a}. The recent study of supernova remnant HESS J1731-347~\cite{LightNS}, hinting to the existence of a low-mass neutron star, is an additional motivation for this search. A discovery would point to a new formation channel, and possibly new physics. An absence of detection improves constraints on dark matter models involving such objects.

During O3, MBTA ran an offline search for CBC events comprising an SSM component (hereafter \enquote{SSM search} for simplicity), along with the PyCBC and GstLAL pipelines~\cite{SSMO3a,SSMO3b}. No events were observed at the time. During the O4 run, we also run an online SSM search, separate from the main search.

\subsection{Template bank}
\begin{figure*}
    \centering
    \includegraphics[width=\textwidth]{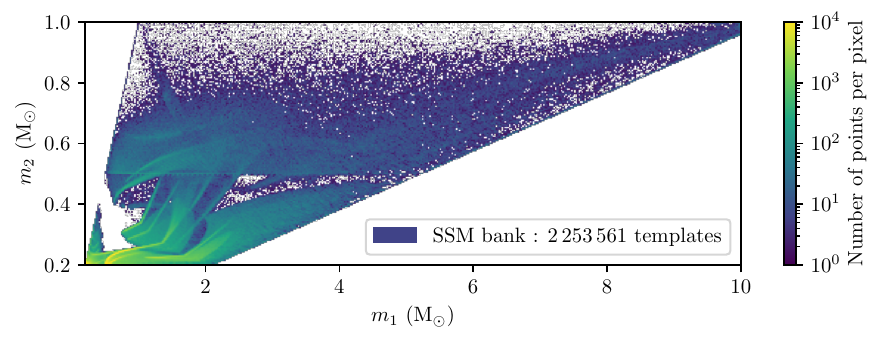}
    \caption{Virtual template bank for the O4a sub-solar mass search in the $(m_1,m_2)$ plane.}
    \label{fig:ssmbank}
\end{figure*}

Similarly to the main search, the SSM search needs three template banks: the virtual template bank and the two real template banks corresponding to the high-frequency and low-frequency bands. The three banks are generated using a geometric algorithm; the parameters can be found in Table ~\ref{table:O4UberBankParam}. The bank contains about 2.2 million virtual templates, which represents an $18\%$ increase compared to the O3 offline SSM bank (due to evolution of the sensitivity curve). The distribution of the templates across the mass dimensions of the parameter space is displayed in Figure~\ref{fig:ssmbank}.

\subsection{Processing}
Most of the pipeline operation is similar to the main search; three important differences exist: 
\begin{itemize}
    \item The templates start at \qty{45}{Hz} in order to reduce the computational cost of the search. This results in a 7\% loss in $\mathrm{SNR}$ compared to starting at \qty{24}{Hz} like the main search
    \item The \snrexcess uses a different max SNR timeseries model, appropriate for the SSM template bank.
    \item Significance estimate: As there is no existing astrophysical prior for an SSM population, we do not compute \pastro for these events, and compute the FAR using the same method as in O3~\cite{MbtaO3}; however, we now also consider single-detector triggers, in a similar fashion to the main search.

\end{itemize}

\subsection{Online}
During the O4a run, we operated the SSM search online for the first time, to be able to respond quickly to a candidate, if any, and to pave the way towards sending public alerts in the future. The configuration used was close to the offline search, with the following differences:
\begin{itemize}
    \item Single-detector triggers were not considered.
    \item As in the online main search, the \enquote{$E_R$} method was used instead of \snrexcess. 
\end{itemize} 
Candidates with FAR values below 2/hour were submitted to the GraceDB database. However, public alerts were not enabled in the low-latency system.

\section{Early-warning search}
\label{sec:early-warning}

The detection of multi-messenger events is one of the important goals of the \lvk: such events, while rare, enable a tremendous amount of science. Only one such event (GW170817~\cite{GW170817orig}) has been observed up until now, but proved extremely useful to the scientific community, providing insights on the behaviour of extreme matter~\cite{GW170817Strontium} and putting tight constraints on the speed of gravity~\cite{GW170817}.

A promising course of action to increase the probability of observing such an event (in particular in channels requiring the instruments to be pointed in the correct direction at the event time) are the so-called \enquote{early-warning} alerts. If the GW signal from the source is strong enough, it is possible to detect it before the merger actually takes place, allowing electromagnetic observatories to quickly react and hopefully observe the immediate aftermath of the merger. This is especially true for gamma-ray bursts, which are typically expected a few seconds after the merger. However, implementing such alerts is not trivial, as a large part of the SNR comes from the last few seconds of the coalescence, and requires important optimizations in order to achieve a significantly lower latency.

\begin{table}
    \centering
    \begin{tabular}{|c|c|c|c|c|} \hline
        $f_{max}$[$\unit{\hertz}$] & 34  & 42 & 50 & 58  \\ \hline
         \multirow{2}{*}{$f_0$[$\unit{\hertz}$]} & \multicolumn{4}{c|}{24 (search)}  \\ 
         & \multicolumn{4}{c|}{15 (bank generation)} \\ \hline
        \makecell{Fraction of\\recovered SNR} & 19\% & 31\% & 44\% & 52\% \\ \hline
        \makecell{Time to\\merger [s]} & [68,15] & [39,8] & [25,5] & [17,4] \\ \hline
        \makecell{Individual\\masses [\Msun]} & \multicolumn{4}{c|}{$m_1,m_2 \in [1,2.5]$} \\ \hline
        Individual spins & \multicolumn{4}{c|}{$s_1^z,s_2^z \in [-0.05,0.05]$} \\ \hline
        Minimal match  & \multicolumn{4}{c|}{$0.98$} \\ \hline
        \makecell{Number of\\templates} & 1562 & 2364 & 4990 & 5045 \\ \hline
        
    \end{tabular}
    \caption{Parameters for the early-warning searches. $f_0$ and $f_{max}$ are defined as in Table~\ref{table:O4UberBankParam}. As for the main search, only aligned spins are taken into account. The SNR fractions are given relative to the \qtyrange[range-units=single,range-phrase=-]{24}{2048}{\hertz} band used by the main search, for the LIGO Hanford detector. The times from $f_{max}$ to merger are the maximum and minimum over the explored parameter space.}
    \label{tab:EWbank}
\end{table}

The early-warning searches~\cite{TheseChristopher} focus on BNS signals, and their principle is identical to the main search, except that the templates are cut at a frequency $f_\mathrm{max}$ much lower than the \qty{2048}{\hertz} for the main search. Table~\ref{tab:EWbank} gives the ranges of achievable time gain for each search for a reference system. Choosing the upper frequency cutoff requires a compromise between the strength of the signals that can be detected (the lower the maximum frequency is, the stronger the signal needs to be) and the latency with which they are reported. We choose to run several instances of the early-warning search with different maximum frequencies in order to maximize the chances of reporting an interesting event; the chosen upper frequency limits for filtering are 42, 50 and \qty{58}{\hertz}, each step leading to roughly a factor 2 increase in collected $\mathrm{SNR}^2$. A \qty{34}{\hertz} search was also implemented, but it was difficult to build a reliable background model due to low-frequency artifacts, leading to this particular search being turned off early during the run. Early-warning searches are performed in a single frequency band for simplicity and due to the small amount of templates. Each of the early-warning searches has a specific template bank, generated using the same geometric algorithm as the BNS seed from the main bank; the parameters are described in Table~\ref{tab:EWbank}. They only cover the BNS parameter space: this represents both the most interesting signals as they are more likely to have an electromagnetic counterpart, and the ones which are the most likely to be detected early as they are longer.

The early-warning searches have other differences with the main search:
\begin{itemize}
    \item The frequency upper limit for computing the gating range is set to \qty{40}{\hertz} (in order to focus on glitches in the frequency range specific to early-warning, such as scattered light~\cite{VirgoDetCharO3}), and uses a \qty{0.3125}{\second} tapering window in order to optimize the latency.
    \item It does not use the SNR Excess feature (instead using the \enquote{$E_R$} method as for the online main search).
    \item We consider only HL coincident events, without filtering Virgo data.
    \item The computation of the FAR and of \pastro is slightly different from the main search. As all templates correspond to BNS signals, we assume that $\pastro = \pbns$; we compute \pastro over a single bin covering the entire parameter space. The foreground rate is rescaled based on an estimate of the percentage of SNR recovered by the early-warning search compared to the main search (given in Table~\ref{tab:EWbank}).
\end{itemize}
  
No significant early-warning events were observed by MBTA during O4a. During the O4b run, we reported two moderately significant candidates (S240429an and S240623dg), which were later retracted as no corresponding events were found by the main search.

\section{Computing resources usage}
\label{sec:computing}
Running expensive computations is a growing concern among scientific collaborations as they strive to reduce the environmental impact of their activities. In the spirit of addressing this issue, we report on the computing resources used by MBTA.

The main benefit of the two-band architecture is a large reduction of the computational cost of the filtering. Table~\ref{tab:resources} details the number of CPUs needed to run an instance of the different search types. As a comparison, the hardware used for all MBTA online operations (including the main search, extra instances of the search for test purposes, monitoring and development servers) represents around $2\%$ of the computing resources allocated to online analyses at the main LVK low-latency computing centers (at CIT and Virgo), or $3.5\%$ of the CPU-based resources allocated to CBC search pipelines.

A rough estimate of the carbon footprint based on~\cite{GreenAlgorithms} shows that around 22.3 tons of $\mathrm{CO}_2$ were emitted during the online O4a run (including all instances of the pipeline). A single offline run of the pipeline emits around \qty{600}{\kilo\gram} of $\mathrm{CO}_2$ (corresponding to $1.89$ M HS06.h\footnote{HS06 or HEP-SPEC06 is a benchmarking procedure commonly used in high-energy physics.}) for the main search and 1.5 tons for the SSM search (note that the large difference compared to the online is mostly due to the lower-carbon electrical mix in France, where the offline run is done, compared to Italy, where the online search is hosted).
\begin{table*}[ht]
    \centering
    \begin{tabular}{l|cccc}
        Search & Main  & SSM & \makecell{Early-warning\\(all)} & \makecell{Development and\\validation} \\ \hline
        Core count & 128 & 160 & 88 & 269\\
        Processor type & Xeon Gold 6242 & Xeon Gold 6242 & Xeon E5-2680 & Mixed \\
        Resource percentage & 20\% & 24\% & 14\% & 42\%  \\
    \end{tabular}
    \caption{Computational resources used by the various MBTA searches during O4a. For each search type, there are two instances, one running on data and the other on data with injections for monitoring purposes.}
    \label{tab:resources}
\end{table*}

\section{Conclusion}
We have described many improvements to the MBTA pipeline, including the new \snrexcess noise rejection method,  a tighter relationship between the FAR and \pastro (which is now computed in low-latency) and the inclusion of single-detector triggers. Two new online searches were also added, with the sub-solar mass being run in low-latency for the first time, and the early-warning search.
The pipeline was able to contribute to many of low-latency alerts during the O4 run, achieving a latency below 30 seconds. Better glitch management led to a smaller number of retracted alerts compared to O3. Further improvements will continue to be implemented while analyzing  new datasets offline and be deployed online in the future.

\section*{Acknowlegements}
We express our gratitude to Gijs Nelemans and Jérôme Novak for their work in carrying out the internal LVK review of the MBTA pipeline. We thank our LVK collaborators from the CBC and low-latency groups for constructive comments. We also thank Dimitri Estevez for commenting on an early version of the paper. We thank Reed Essick (and the entire LVK Rates and Populations group) for developing the foreground model and injection sets used within our work. This analysis exploits the resources of the computing facility at the EGO-Virgo site, and of the Computing Center of the Institut National de Physique Nucléaire et Physique des Particules (CC-IN2P3 / CNRS). This research has made use of data or software obtained from the Gravitational Wave Open Science Center (gwosc.org), a service of the LIGO Scientific Collaboration, the Virgo Collaboration, and KAGRA. This material is based upon work supported by NSF's LIGO Laboratory which is a major facility fully funded by the National Science Foundation, as well as the Science and Technology Facilities Council (STFC) of the United Kingdom, the Max-Planck-Society (MPS), and the State of Niedersachsen/Germany for support of the construction of Advanced LIGO and construction and operation of the GEO600 detector. Additional support for Advanced LIGO was provided by the Australian Research Council. Virgo is funded, through the European Gravitational Observatory (EGO), by the French Centre National de Recherche Scientifique (CNRS), the Italian Istituto Nazionale di Fisica Nucleare (INFN) and the Dutch Nikhef, with contributions by institutions from Belgium, Germany, Greece, Hungary, Ireland, Japan, Monaco, Poland, Portugal, Spain. KAGRA is supported by Ministry of Education, Culture, Sports, Science and Technology (MEXT), Japan Society for the Promotion of Science (JSPS) in Japan; National Research Foundation (NRF) and Ministry of Science and ICT (MSIT) in Korea; Academia Sinica (AS) and National Science and Technology Council (NSTC) in Taiwan.

\appendix
\renewcommand{\thesection}{Appendix \Alph{section}}
\sectionfont{\fontsize{11}{15}\selectfont}
\section{Multiple FAR(\pastro) models} 
\label{sec:farpastroappendix}

\begin{figure}
    \centering
    \includegraphics[width=0.5\textwidth]{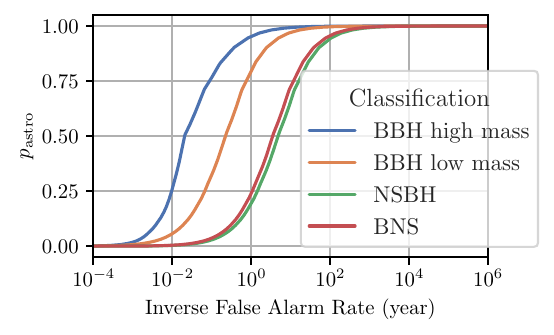}
    \caption{FAR(\pastro) parametrizations for the online analysis, using one model per source type.}
    \label{fig:farpastroseveral}
\end{figure}

In section~\ref{sec:FAR}, we detailed how we obtain a global FAR(\pastro) parametrization for the offline run. For the online configuration, we choose a slightly different approach in order to increase the chances of detecting a multi-messenger event. Instead of summing the $\text{FAR}_i(\pastro)$ over all bins and therefore all source types, we build four different models corresponding to different source types: BNS, NSBH, low-mass BBH (BBH-low) and high-mass BBH (BBH-high), with a split at $\mchirp > 7 \Msun$ for the BBH motivated by the threshold for single-detector triggers. This is achieved by modifying equation~\ref{eq:FAR_global} as:
\begin{equation}
    \mathrm{FAR}_s(\pastro) = w_s \sum_{i\in \mathcal{O}_s} \mathrm{FAR}_i(\pastro)
\end{equation}
where $\mathcal{O}_s$ is the set of bins for which the source $s$ is the dominant one (i.e. represents the largest number of injections found in this bin when building the \pastro model). When associating a FAR to a candidate, we use the model corresponding to its dominant source type. Effectively, this divides the parameter space into four different \enquote{searches}. We thus need to apply a trial factor $w_s$ to each of the models. We choose here to share the background budget equally between BNS, NSBH and BBH, giving a trial factor of $1/3$ for each (for the BBH, this is further split with BBH-low representing 10\% and BBH-high the remaining 90\%, determined empirically with the \mchirp distribution of the population model). This gives values of $w_\mathrm{BNS} = w_\mathrm{NSBH} = 0.333$, $w_\mathrm{BBH-low} = 0.033$ and $w_\mathrm{BBH-high} = 0.3$. By placing the BNS and NSBH on an equal footing with the BBH, we thus encourage their detection. Note that this process only affects the FAR value; the \pastro is unchanged. Figure~\ref{fig:farpastroseveral} shows these different models: for a same \pastro value, the BNS and NSBH will be assigned a higher IFAR value (meaning the event is given more significance) than the BBH.

\bibliographystyle{style}
\bibliography{bibliography.bib} 
\end{document}